\documentclass[a4paper,superscriptaddress,preprint]{revtex4-1}
\usepackage[pdftex]{graphicx}
\pdfoutput=1
\usepackage[]{amsmath}
\usepackage{sidecap}
\usepackage{hyperref}

\begin{document}
\title{Space-Time Area in Atom Interferometry}

\author{G. D. McDonald}
\email{gordon.mcdonald@anu.edu.au}
\homepage{http://atomlaser.anu.edu.au/}
\author{C. C. N. Kuhn}

\affiliation{Quantum Sensors Lab, Department of Quantum Science, Australian National University, Canberra, 0200, Australia}

\date{\today} 

\begin{abstract}
It is a commonly stated that the acceleration sensitivity of an atom interferometer is proportional to the space-time area enclosed between the two interfering arms \cite{ClaudeSTA, PhysRevLett.108.090402,DebsBECgrav}. Here we derive the interferometric phase shift for an extensive class of interferometers, and explore the circumstances in which only the inertial terms contribute. We then analyse various configurations in light of this geometric interpretation of the interferometric phase shift.
\end{abstract}

\maketitle

\section{Atom-light interactions}
\label{sec:SeparatedAITheory-PhaseShift}

\begin{figure}[h]
\centering{}
 \includegraphics[width=1\columnwidth]{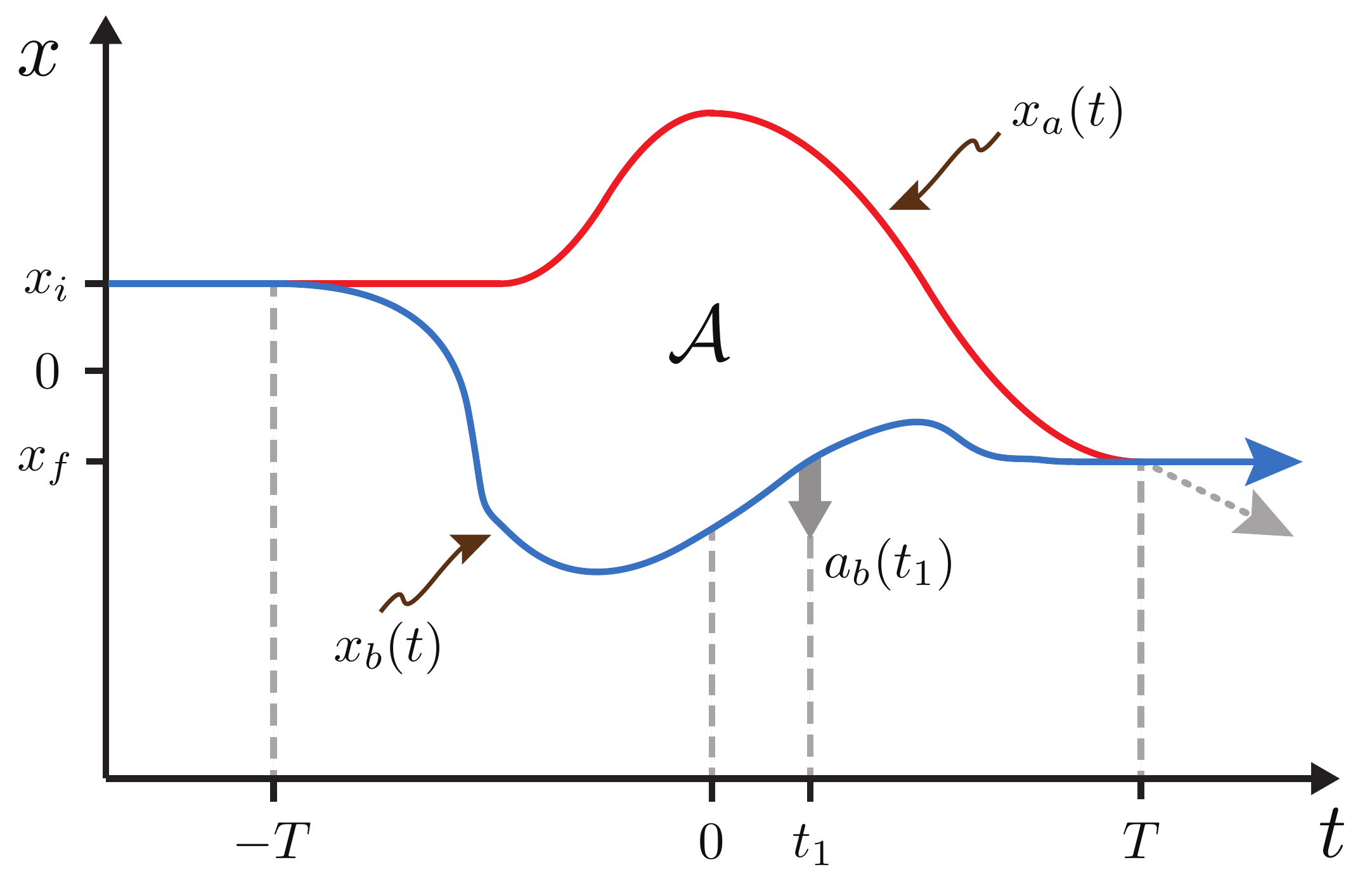}
 \caption{(Color online) The two paths $a$ and $b$ form an arbitrary closed interferometer. For clarity, the diagram is drawn with $\mathbf{g}=0$. We are calculating the difference in phase accumulated along two paths which exit the interferometer at the same place and with the same final velocity, here shown as a blue arrow. The alternate exit path is shown as a grey dotted arrow. The area enclosed between the two paths in this space-time diagram is the space-time area, $\mathcal{A}$.}
 \label{SquiggleInt}
 \end{figure}

If a particle experiences a linearly varying potential term $V(\mathbf{x})=m\mathbf{g}\cdot\mathbf{x}$ and is otherwise free, it will experience an acceleration $\mathbf{g}$. The two arms of the interferometer, labelled $a$ and $b$, experience additional accelerations $\mathbf{\tilde{a}}_a(t)$ and $\mathbf{\tilde{a}}_b(t)$ respectively at time $t$, which we shall define from $t=0$ in the middle of our interferometer as in Fig. \ref{SquiggleInt}. These may comprise both the inertial acceleration $\mathbf{g}$ and any other acceleration from time-varying potentials used to generate the interferometer (e.g. Bragg diffraction pulses, Bloch lattice accelerations, magnetic field gradients etc.). 

\begin{SCfigure}
\centering{}
 \includegraphics[width=0.3\columnwidth]{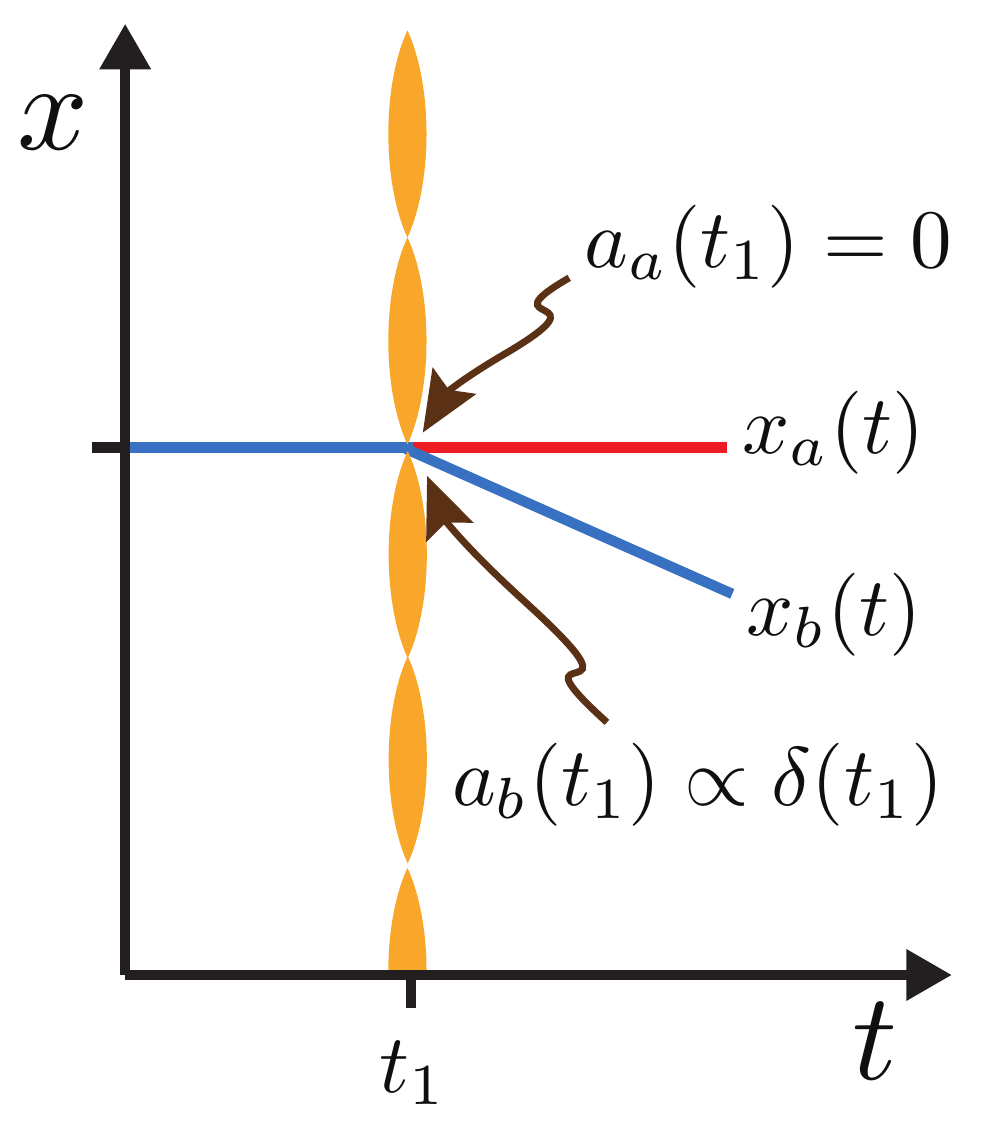}
 \caption{Treat each Bragg diffraction as a $\delta$-function acceleration kick on the diffracted path, which in this figure is path $b$. The other path (path $a$) experiences no effect.  }
 \label{BraggDiff}
 \end{SCfigure}

For example, to treat a Bragg diffraction pulse at time $t_1$ which imparts a downwards velocity kick of $2n\hbar \mathbf{k}/m$ to path $b$, (see Figure \ref{BraggDiff}) we can write the kicked path's acceleration in the inertial frame as $\mathbf{\tilde{a}}_b(t)=-\delta(t-t_1)\frac{2n\hbar \mathbf{k}}{m}$, and the acceleration of the unkicked path (also in the inertial, freely-falling frame) as $\mathbf{\tilde{a}}_a(t)=0$. For a more complicated sequence of kicks, we can simply take the sum of each one.

As another example, constant acceleration in an optical Bloch lattice can be expressed as a classical acceleration $\mathbf{\tilde{a}}_b(t) =  \frac{2n_{b}\hbar \mathbf{k}}{m\tau_b}$ where the number of Bloch oscillations is given as $n_b$ and the time for a single oscillation is $\tau_b$. Alternatively, it may be expressed as one $2\hbar\mathbf{k}$ kick every Bloch oscillation period $\tau_b$, starting $\tau_b/2$ from the beginning, e.g. $\mathbf{\tilde{a}}_b(t) =  \sum_{i=1}^{n_b}\frac{2\hbar \mathbf{k}}{m}\delta(t-t_i)$ where $t_i=t_0+\tau_b(i+1/2)$ and $t_0$ is start time of the Bloch lattice acceleration. Either treatment gives the same space-time area in the total interferometer. For such a treatment to correctly the phase shift along each path individually the atoms must experience a whole number of Bloch oscillations. A small correction when this is not the case has been investigated in Ref.~\cite{KasevichBlochAnalysis}.

\section{Interferometer Phase Shift}
\label{sec:AITheory-PhaseShift}

\subsection{Action}
The phase shift of an atom interferometer can be calculated by the path integral 
formalism. Consider the classical action $S_a$ of a particle of mass $m$ moving 
along a path $\mathbf{\mathbf{x}_a}(t)$ with velocity $\mathbf{v}_a(t)$ and experiencing an acceleration 
$\mathbf{a}_a(t)$,

\begin{eqnarray}
S_a &=& \int_{-T}^{T}K_a-V_a\,\, dt
\\&=& m\int_{-T}^{T}\frac{{v_a^2}}{2}+\mathbf{a}_a\cdot\mathbf{x}_a\,\,ds.
\end{eqnarray}

where we have defined the potential term $V_a(t)=-m\mathbf{a}_a(t)\cdot\mathbf{x}_a(t)$. The phase shift of an interferometer consisting of arms traversing the classical paths $a$ and $b$ is given by

\begin{eqnarray}
\Delta \Phi &=& \frac{\Delta S}{\hbar} 
\\  &=& \frac{S_a-S_b}{\hbar}
\\  &=& \frac{m}{\hbar}\int_{-T}^{T} \frac{{v_a^2-v_b^2}}{2}+\mathbf{a}_a\cdot\mathbf{x}_a-\mathbf{a}_b\cdot\mathbf{x}_b\, ds
\\  &=& \frac{m}{\hbar}\int_{-T}^{T} \frac{{\Delta \left[v^2\right]}}{2}+\Delta \left[\mathbf{a}\cdot\mathbf{x}\right]\, ds.
\label{eq:simplestart}
\end{eqnarray}

If we now make the distinction between the acceleration  $\mathbf{g}$ the atoms would experience inertially, in absence of the interferometric sequence and the acceleration $\mathbf{\tilde{a}}_i$ the atoms feel along path $i$ specifically because of the sequence, along with the corresponding separations for velocity and position,

\begin{align}
\mathbf{a}_i&=\mathbf{\tilde{a}}_i+\mathbf{g}\\
\mathbf{v}_i&=\mathbf{\tilde{v}}_i+\mathbf{v}_g\\
\mathbf{x}_i&=\mathbf{\tilde{x}}_i+\mathbf{x}_g
\end{align}
we can expand the kinetic energy term in the integrand of equation (\ref{eq:simplestart}) to be
\begin{eqnarray}
\frac{{\Delta \left[v^2\right]}}{2} &=& \frac{\mathbf{\tilde{v}}_a^2-\mathbf{\tilde{v}}_b^2}{2}+\mathbf{v}_g(\mathbf{\tilde{v}}_a-\mathbf{\tilde{v}}_b)
\label{eq:expandedphasekin}
\end{eqnarray}
while the potential term in the integrand of equation (\ref{eq:simplestart}) becomes
\begin{eqnarray}
\Delta \left[\mathbf{a}\cdot\mathbf{x}\right] &=& \mathbf{\tilde{a}}_a\cdot\mathbf{\tilde{x}}_a-\mathbf{\tilde{a}}_b\cdot\mathbf{\tilde{x}}_b+(\mathbf{\tilde{a}}_a-\mathbf{\tilde{a}}_b)\cdot\mathbf{\tilde{x}}_g+\mathbf{g}\cdot\Delta\mathbf{\tilde{x}}
\label{eq:expandedphasepot}
\end{eqnarray}

If we consider just the integral of the potential term, and integrate by parts

\begin{eqnarray}
&&\int_{-T}^{T} \Delta \left[\mathbf{a}\cdot\mathbf{x}\right]\, dt
\\  &=&\int_{-T}^{T}  \mathbf{\tilde{a}}_a\cdot\mathbf{\tilde{x}}_a-\mathbf{\tilde{a}}_b\cdot\mathbf{\tilde{x}}_b
+(\mathbf{\tilde{a}}_a-\mathbf{\tilde{a}}_b)\cdot\mathbf{x}_g
+\mathbf{g}\cdot\Delta\mathbf{\tilde{x}}\, dt
\\  &=& {\left[\mathbf{\tilde{v}}_a\cdot\mathbf{x}_a-\mathbf{\tilde{v}}_b\cdot\mathbf{x}_b \right]}_{-T}^{T}
-\int_{-T}^{T} \tilde{v}_a^2-\tilde{v}_b^2\, dt\nonumber\\
&&\qquad-\int_{-T}^{T} \mathbf{v}_g(\mathbf{\tilde{v}}_a-\mathbf{\tilde{v}}_b)\, dt+\int_{-T}^{T} \mathbf{g}\cdot\Delta\mathbf{\tilde{x}}\,dt
\end{eqnarray}

then Eq.(\ref{eq:simplestart}) becomes

\begin{eqnarray}
\Delta \Phi &=& \frac{m}{\hbar}\int_{-T}^{T} \frac{{\Delta \left[v^2\right]}}{2}+\Delta \left[\mathbf{a}\cdot\mathbf{x}\right]\, ds.\\
&=& \frac{m}{\hbar}\left(\left[\mathbf{\tilde{v}}_a\cdot\mathbf{x}_a-\mathbf{\tilde{v}}_b\cdot\mathbf{x}_b \right]_{-T}^{T}-\frac{1}{2}\int_{-T}^{T} \tilde{v}_a^2-\tilde{v}_b^2\, dt+\int_{-T}^{T} \mathbf{g}\cdot\Delta\mathbf{\tilde{x}}\,dt\right)
\label{eq:simplenext}
\end{eqnarray}

Let us call the boundary term $\Delta \Phi_{\text{sep}}$, the separation phase
\begin{align}
\Delta \Phi_{\text{sep}}=&\frac{m}{\hbar}\left[\mathbf{\tilde{v}}_a\cdot\mathbf{x}_a-\mathbf{\tilde{v}}_b\cdot\mathbf{x}_b \right]_{-T}^{T}\\
=&\left[\mathbf{\tilde{k}}_a\cdot\mathbf{x}_a-\mathbf{\tilde{k}}_b\cdot\mathbf{x}_b \right]_{-T}^{T}\\
=&\mathbf{\tilde{k}}_a(T)\cdot\mathbf{x}_a(T)-\mathbf{\tilde{k}}_b(T)\cdot\mathbf{x}_b(T)-\mathbf{\tilde{k}}_a(-T)\cdot\mathbf{x}_a(-T)+\mathbf{\tilde{k}}_b(-T)\cdot\mathbf{x}_b(-T)
\end{align}
For the final states to interfere they must have the same final position $\mathbf{x}_b(T)=\mathbf{x}_a(T)$, and for this interference to persist in the far-field they must have the same final velocity, $\mathbf{\tilde{k}}_b(T)=\mathbf{\tilde{k}}_a(T)$ so the separation phase depends only upon the initial states,
\begin{align}
\Delta \Phi_{\text{sep}}=\mathbf{\tilde{k}}_b(-T)\cdot\mathbf{x}_b(-T)-\mathbf{\tilde{k}}_a(-T)\cdot\mathbf{x}_a(-T)\,.
\end{align}
In the case that the initial velocities are the same then the separation phase becomes
\begin{align}
\Delta \Phi_{\text{sep}}&=-\mathbf{\tilde{k}}(-T)\cdot\Delta\mathbf{x}(-T)\\
&=-\mathbf{\tilde{k}}_i\cdot\Delta\mathbf{x}_i\,.
\end{align}
and if the initial separation $\Delta\mathbf{x}_i=0$ (i.e. the interferometer is closed) then there is no contribution from the separation phase.

The kinetic term can be re-written in terms of frequency, through the relation $\tilde{E}_\text{kin}=\hbar\tilde{\omega}=\frac{\hbar^2\tilde{k}^2}{2m}$
\begin{align}
\Delta\Phi_{\text{kin}}&=-\frac{m}{2\hbar}\int_{-T}^{T} \tilde{v}_a^2-\tilde{v}_b^2\, dt\\
&=\int_{-T}^{T}\tilde{\omega}_b-\tilde{\omega}_a\,dt
\end{align}
and this term has been used for measurements of the recoil frequency e.g. in Refs \cite{PhysRevLett.103.050402,HMullerBlochBraggBloch,ContrastInterferometry}.

So the total interferometer phase shift becomes

\begin{eqnarray}
\Delta \Phi &=&-\mathbf{\tilde{k}}_i\cdot\Delta\mathbf{x}_i+\int_{-T}^{T} \tilde{\omega}_b-\tilde{\omega}_a\, dt+\frac{m}{\hbar}\int_{-T}^{T} \mathbf{g}\cdot\Delta\mathbf{\tilde{x}}\,dt\\
&=&\quad\Delta\Phi_{\text{sep}}\quad+\qquad\Delta\Phi_{\text{kin}}\qquad+\qquad\Delta\Phi_{\text{inertial}}
\label{eq:simplelast}
\end{eqnarray}

There are many ways (thorough various symmetries) in which to make ${\Delta\Phi_{\text{kin}}=0}$, which we will discuss in the next section. In any of these cases, for a closed interferometer, the phase shift simplifies to

\begin{eqnarray}
\Delta \Phi &=&\qquad\Delta\Phi_{\text{inertial}}\\
&=&\frac{m}{\hbar}\int_{-T}^{T} \mathbf{g}\cdot\Delta\mathbf{\tilde{x}}\,dt
\label{eq:simplefinal}
\end{eqnarray}
and for a constant acceleration $\mathbf{g}$ this can be pulled out of the integral,
\begin{eqnarray}
\Delta \Phi &=&\frac{m}{\hbar}\mathbf{g}\cdot\int_{-T}^{T} \Delta\mathbf{\tilde{x}}\,dt\\
&=&\frac{m}{\hbar}\mathbf{g}\cdot\mathcal{A}
\label{eq:simplefinal2}
\end{eqnarray}
where we have defined the space-time area on the last line, $\mathcal{A}\equiv\int_{-T}^{T} \Delta\mathbf{\tilde{x}}\,dt$. This term is used for measurements of gravity e.g. in Refs. \cite{BestAtomicGravimeter,OurGravimeter,PhysRevA.81.013617,DebsBECgrav,SchmiedmayerBECMZI,PhysRevA.88.043610,aip101}, and is proposed to be used to measure the effect of gravity on antimatter \cite{PhysRevLett.112.121102}. It is also used in the form of Eq. (\ref{eq:simplefinal}) to measure the recoil frequency by applying an additional inertial acceleration in Refs. \cite{Alpha2008,Alpha2011}.

To include the possibility of different internal magnetic states of the atoms, an additional phase shift $\Delta\Phi_{\text{mag}}$ must be added. As the potential energy of a magnetic dipole with dipole moment $\mu$ in a magnetic field is given by $U=-\boldsymbol\mu\cdot\mathbf{B}$, the phase shift is given by

\begin{align}
\Delta\Phi_{\text{mag}}&=-\int_{-T}^{T}\Delta U \,dt\\
&=\int_{-T}^{T}\mathbf{B}\cdot\Delta \boldsymbol\mu \,dt
\label{eq:magshift}
\end{align}

This will apply to Raman interferometry in particular, whereas in Bragg interferometry the atoms stay in the same internal state and so this shift is zero. 

\subsection{Time Symmetries}

There are many arbitrary ways to create an interferometer in which $\Delta\Phi_{\text{kin}}=0$. One way is to enforce either of the following velocity symmetries

\begin{align}
\mathbf{\tilde{v}}_a(t)&=\mathbf{\tilde{v}}_b(-t)\qquad(i)\label{eq:Vsym}\\
&\text{or}\nonumber \\
\mathbf{\tilde{v}}_a(t)&=-\mathbf{\tilde{v}}_b(-t)\qquad(ii)\label{eq:Vantisym}
\end{align}

from which it follows that

\begin{align}
\mathbf{\tilde{v}}_a^2(t)&=\mathbf{\tilde{v}}_b^2(-t)\\
\int_{-T}^{T}\mathbf{\tilde{v}}_a^2(t)\,dt&=\int_{-T}^{T}\mathbf{\tilde{v}}_b^2(-t)\,dt\\
\int_{-T}^{T}\mathbf{\tilde{v}}_a^2(t)-\mathbf{\tilde{v}}_b^2(t)\,dt&=0\\
\Delta\Phi_{\text{kin}}=0
\end{align}

These symmetries also help to cancel other shifts in real interferometers such as stark shifts and others as described in section \ref{sec:potenergyshifts}.
\subsection{Laser Phase}

Any laser interaction which kicks a path in such a way to increase the space-time area of the interferometer, its phase should be added, and any interaction which kicks a path in such a way as to decrease the interferometer's space-time area, its phase should be subtracted. Thus

\begin{eqnarray}
\phi_L &=& \sum_{\text{increases }\mathcal{A}}\phi_i\quad - \quad\sum_{\text{decreases }\mathcal{A}}\phi_i
\end{eqnarray}

where  each $\phi_i$ is the laser phase accumulated along a certain path when that path experiences a 2-photon-recoil change in momentum. This is a different way to state the same result as in Ref. \cite{BestAtomicGravimeter}, but simpler as we are dealing with Bragg and not Raman transitions and so do not have to deal with changes in internal state.

\section{Extensions}

\subsection{Constant energy offset}
\label{sec:potenergyshifts}

If one of the trajectories experiences a spatially constant potential energy offset $V_0(t)$, for example due to a state-dependent stark shift, this will cause a phase offset in the interferometer of $\frac{1}{\hbar}\int V_0(t)\, dt$. In the case where this shift is cancelled by an equal shift in the opposite direction later in time, there is zero net phase shift. This cancellation occurs in particular in the constant-acceleration Bloch interferometer configuration, because the un-accelerated arm of the interferometer experiences a stark shift due to the average optical lattice intensity.

%\subsection{Coriolis effect}
%
%To include the Coriolis effect as well as gravity we may very well have stated the symmetry requirement as 
%\begin{equation}
%	\frac{\mathbf{a}_b(t)+\mathbf{a}_a(-t)}{2}=\mathbf{g}+2\mathbf{\Omega}\times\mathbf{v}\,.
%\end{equation}
%in which case we simply replace $\mathbf{g}$ in the phase shift;
%
%\begin{eqnarray}
%\Delta \Phi &=& \frac{m\left[\mathbf{g}+2\mathbf{\Omega}\times\mathbf{v_y}\right]}{\hbar}\cdot \mathbf{\mathcal{A}}
%\end{eqnarray}

\subsection{Vibrations and time-varying $\mathbf{g}$}
\label{sec:vib}

There is nothing in the derivation above preventing $\mathbf{g}$ being considered an arbitrary function of time. In this case it must say within the integral
\begin{eqnarray}
\Delta \Phi_{\text{inertial}}&=& \frac{m}{\hbar} \int_{-T}^{T} \mathbf{g}\cdot\Delta \mathbf{x}\, dt\,.
\end{eqnarray}
 A symmetric variation of the form $\mathbf{g}(t)=\mathbf{g}(-t)$ is used in measurements of the fine structure constant, e.g. Refs \cite{Alpha2011,Alpha2008}. After a Ramsey-Bord\'e configuration interferometer has had its paths separated, \emph{both} arms are loaded into the same Bloch lattice and accelerated, then decelerated again, effectively changing $\mathbf{g}$ symmetrically in a way proportional to the recoil frequency $\omega_{rec}=\frac{\hbar k^2}{2m}$.

We can also consider the effect of a vibration and how this will couple in to our interferometer signal. We can write a sinusoidal acceleration with frequency $\omega$ as

\begin{eqnarray}
\mathbf{g}=\mathbf{a}_c \cos(\omega t)+\mathbf{a}_s\sin(\omega t)
\end{eqnarray}

%and consider the effect of the perturbational Lagrangian $m\mathbf{a}\cdot\mathbf{x}$ on the interferometric phase of the \emph{unperturbed} path difference $\Delta \mathbf{x}$\footnote{This makes intuitive sense if we consider that it is not the path that is changing, but only the phase of the applied optical lattice, due to vibrations of the fibre-couplers etc.}:
which will cause a phase shift in the interferometer output of
\begin{align}
\Delta\Phi_{\text{inertial}}&= \frac{m}{\hbar} \int_{-T}^{T} \left[\mathbf{a}_c \cos(\omega t)+\mathbf{a}_s\sin(\omega t)\right]\cdot\Delta \mathbf{\tilde{x}}\,dt\\
&= \frac{m}{\hbar}\left[\mathbf{a}_c \cdot \int_{-T}^{T} \cos(\omega t)\Delta \mathbf{\tilde{x}}\,dt+\mathbf{a}_s\cdot\int_{-T}^{T} \sin(\omega t)\Delta \mathbf{\tilde{x}}\,dt\right]\\
&=\frac{m}{\hbar}\left[\mathbf{a}_c \cdot \mathcal{A}_c(\omega)+\mathbf{a}_s\cdot \mathcal{A}_s(\omega)\right]
\end{align}
where on the last line we have defined the effective space-time areas $ \mathcal{A}_c(\omega)$ and $ \mathcal{A}_s(\omega)$ for a given frequency of vibration $\omega$. Note that the space-time area as defined before is $\mathcal{A}=\mathcal{A}_c(0)$.

If the path separation is symmetric about $t=0$, i.e. $\Delta \mathbf{\tilde{x}}(t)=\Delta \mathbf{\tilde{x}}(-t)$ then the sine term cancels $ \mathcal{A}_s(\omega)=0$, leaving only the cosine term $ \mathcal{A}_c(\omega)$. A useful measure of vibration sensitivity when this symmetry is present is the relative acceleration sensitivity as a function of frequency, normalised to the sensitivity at $\omega=0$. This is given by $\mathcal{R}(\omega)\equiv\frac{|\mathcal{A}_c(\omega)|}{|\mathcal{A}|}$. From this unitless ratio we can deduce the phase response to an acceleration with frequency $\omega$ via

\begin{align}
\Delta\Phi_{\text{inertial}}=\frac{m\mathcal{R}(\omega)}{\hbar}\mathbf{a}_c\cdot\mathcal{A}
\label{eq:VibSensCos}
\end{align}

If the path separation is antisymmetric about $t=0$, i.e. $\Delta \mathbf{\tilde{x}}(t)=-\Delta \mathbf{\tilde{x}}(-t)$ then the cosine term cancels $ \mathcal{A}_c(\omega)=0$, leaving only the sine term $ \mathcal{A}_s(\omega)$. Therefore, these types of interferometer are insensitive to a constant acceleration. A useful measure of vibration sensitivity when this symmetry is present is the relative acceleration sensitivity as a function of frequency, normalised to the sensitivity to a constant acceleration of the corresponding symmetrized interferometer, i.e. $\mathcal{A^*}\equiv\int_{-T}^{T}\left| \Delta\mathbf{\tilde{x}}\right|\,dt$ at $\omega=0$. This is given by $\mathcal{R^*}(\omega)\equiv\frac{|\mathcal{A}_s(\omega)|}{|\mathcal{A^*}|}$. From this unitless ratio we can deduce the phase response to an acceleration with frequency $\omega$ via

\begin{align}
\Delta\Phi_{\text{inertial}}=\frac{m\mathcal{R^*}(\omega)}{\hbar}\mathbf{a}_s\cdot\mathcal{A^*}
\label{eq:VibSensSin}
\end{align}

\subsection{Fourier series decomposition}

If $\mathbf{g}(t)$ can be considered an arbitrary piecewise continuous function of time between $-T\leq t\leq T$, then it can be written as a Fourier series
\begin{align}
\mathbf{g}(t)=\sum_{j=0}^{\infty} \left[ \mathbf{a}_{c,j} \cos\left(\frac{j\pi t}{T}\right)+\mathbf{a}_{s,j}\sin\left(\frac{j\pi t}{T}\right) \right]
\end{align}
for the Fourier coefficients
\begin{align}
\mathbf{a}_{c,j} =\frac{1}{T}\int_{-T}^{T}\cos\left(\frac{j\pi t}{T}\right)\mathbf{a}(t)\,dt
\end{align}
and
\begin{align}
\mathbf{a}_{s,j} =\frac{1}{T}\int_{-T}^{T}\sin\left(\frac{j\pi t}{T}\right)\mathbf{a}(t)\,dt\,.
\end{align}
We can see that the effect of such an arbitrary acceleration will be a phase shift of 
\begin{align}
\Delta\Phi_{\text{inertial}}&= \frac{m}{\hbar}\sum_{j=0}^{\infty}\left[\mathbf{a}_{c,j}\cdot\mathcal{A}_c\left(\frac{j\pi}{T}\right)+\mathbf{a}_{s,j}\cdot\mathcal{A}_s\left(\frac{j\pi}{T}\right)\right]\,.
\end{align}

\subsection{Coriolis Effect}

Similarly to the previous section the common inertial acceleration $\mathbf{g}$ should be kept inside the integral

\begin{eqnarray}
\Delta \Phi &=& \frac{m}{\hbar} \int_{-T}^{T} \mathbf{g}\cdot\Delta \mathbf{x}\, dt
\end{eqnarray}

whereupon the substitution $\mathbf{g}\rightarrow\mathbf{g}+\Omega\times\mathbf{v}$ can be used to perturbatively incorporate the effect of rotation on the interferometer for small constant $\Omega$, e.g. the rotation of the earth \footnote{Technically, this is evaluating a perturbative Lagrangian $\mathcal{L}\approx m\Omega\cdot\mathbf{r}\times\mathbf{v}$ along the unperturbed path, as in Ref. \cite{ClaudeSTA}, and ignoring the term proportional to $\Omega^2$.}. Thus

\begin{eqnarray}
\Delta \Phi_{\text{inertial}} &=& \frac{m}{\hbar} \int_{-T}^{T} \left(\mathbf{g}+\Omega\times\mathbf{v}\right)\cdot\Delta \mathbf{x}\, dt\\
&=&\frac{m}{\hbar} \left(\mathbf{g}\cdot\mathcal{A}-\int_{-T}^{T} \Omega\cdot\Delta \mathbf{x}\times\mathbf{v}\, dt\right)\\
&=&\frac{m}{\hbar} \left(\mathbf{g}\cdot\mathcal{A}-\Omega\cdot\int_{-T}^{T} \Delta \mathbf{x}\times d\mathbf{x}\right)\\
&=&\frac{m}{\hbar} \left(\mathbf{g}\cdot\mathcal{A}-\Omega\cdot\oint \mathbf{x}\times d\mathbf{x}\right)\\
&=&\frac{m}{\hbar} \left(\mathbf{g}\cdot\mathcal{A}-2 \Omega\cdot\mathbf{A}\right)
\end{eqnarray}
which reproduces the well-known Sagnac phase shift as the term on the right, where $\mathbf{A}$ is the vector-area enclosed by the interferometer paths. Under the assumption that all the $\mathbf{k}$-vectors are parallel, the area arises both from an initial velocity $\mathbf{v}_i$ and the mean velocity of the accelerating atoms $\mathbf{g}$, the area $\mathbf{A}$ is given by

\begin{eqnarray}
\mathbf{A}&=& -\mathbf{v}_i\times\mathcal{A}-\mathbf{g}\times\int_{-T}^{T} t\Delta \mathbf{x}\, dt
\end{eqnarray}

where the second term (which goes as a higher power of interferometer time $T$ than the first term) will disappear if $\mathbf{g}\times\mathbf{k}=0$, or if the separation $\Delta\mathbf{x}$ is symmetric about $t=0$. In this case the interferometer phase becomes 

\begin{eqnarray}
\Delta \Phi&=&\frac{m}{\hbar} \left(\mathbf{g}-2 \Omega\times\mathbf{v}_i\right)\cdot\mathcal{A}
\end{eqnarray}

This rotationally-sensitive term is measured in Refs. \cite{PhysRevLett.108.090402,PhysRevLett.111.083001}

%\subsection{General Rotating frame}
%
%In general, the phase 

\subsection{Separation Phase}

The results above all apply to a closed-loop interferometer configuration. If for some reason, the parts of each state which overlap at the end of the interferometer did not originate from the same place at the beginning of the interferometer (for example due to a slight timing offset of the last pulse), then the separation phase is non-zero and we have
  \begin{figure}
\centering{}
 \includegraphics[width=0.9\columnwidth]{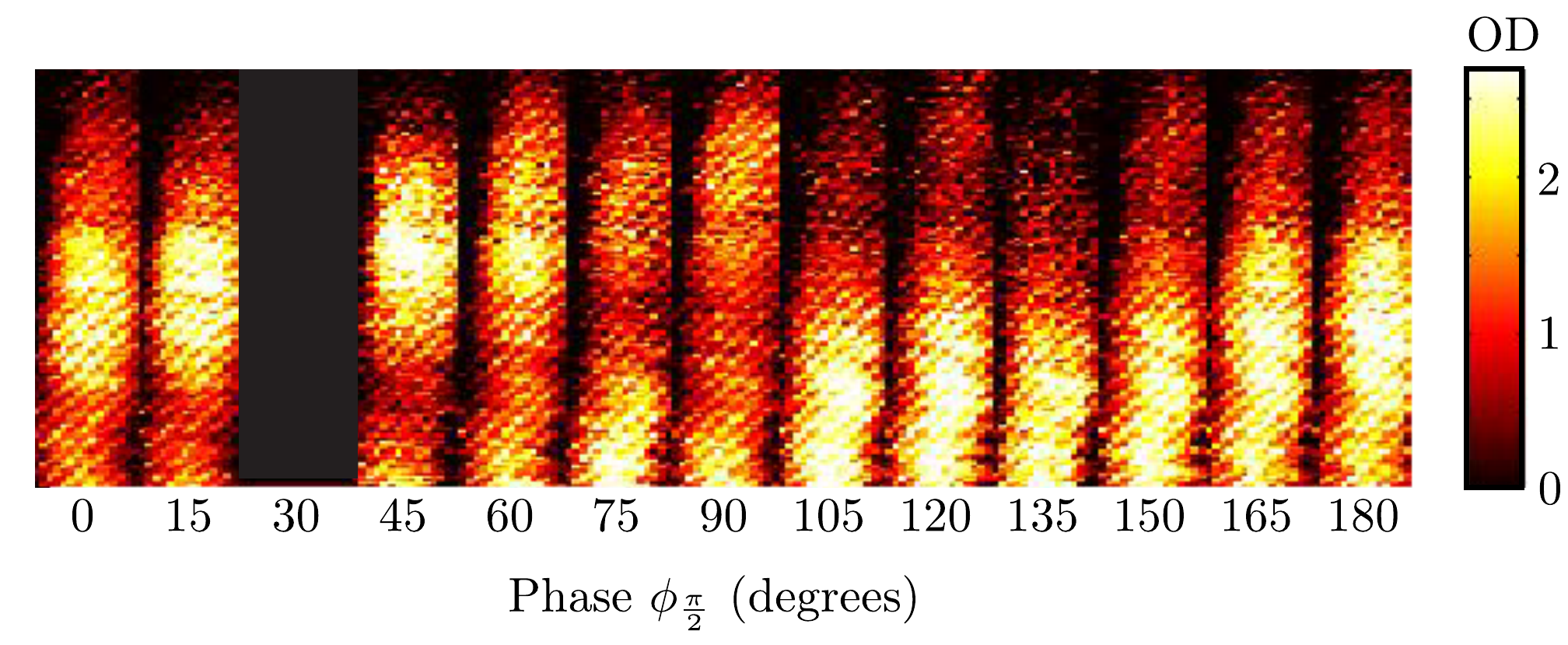}
 \caption{Demonstration of spatial fringes via the separation phase $\Delta\Phi_{\text{sep}}$, performed in our lab. In a Mach-Zehnder interferometer with a small time offset $\delta T$ on the last pulse, this is the $4\hbar k$ output state only, as the interferometric phase $\phi_{\frac{\pi}{2}}$ of the recombination $\frac{\pi}{2}$ pulse is varied. In the image for $\phi_{\frac{\pi}{2}}=30^o$ the DDS generated the pulse sequence incorrectly so this image is not shown.}
 \label{fig:SpatialFringe}
 \end{figure} 
\begin{eqnarray}
\Delta \Phi_{\text{sep}}&=&\mathbf{k}_{e}\cdot\Delta \mathbf{x}_i
\end{eqnarray}
where $\Delta \mathbf{x}_i$ is the initial separation of the two finally overlapped endpoints, and {${\mathbf{k}_e=2n\mathbf{k}}$} is the initial momentum separation, i.e. the momentum separation of the \emph{first} bragg kick. If this is due to a timing offset $\delta T$ then the separation phase is given by
\begin{eqnarray}
\Delta \Phi_{\text{sep}}&=&\mathbf{k}_{e}\cdot\frac{\hbar \mathbf{k}_e }{m}\delta T\\
&=&8 n^2\omega_{\text{rec}}\,\delta T
\end{eqnarray}
where $\omega_{\text{rec}}=\frac{\hbar k^2}{2m}$ is the single-photon recoil frequency. Likewise, there is a momentum separation phase for the case in which different final momentum states did not originate from the same initial momentum state, for instance due to the spread of momentum across a cold atom cloud, in an interferometer with a time delay $\delta T$. This is given by

\begin{eqnarray}
\Delta \Phi_{\text{sep}}&=\mathbf{x}_i\cdot\Delta\mathbf{k}
\end{eqnarray}
and is experimentally demonstrated in Fig. \ref{fig:SpatialFringe}.

\section{Examples}

\subsection{Mach-Zehnder (MZ)}

\begin{figure}[h]
\centering{}
 \includegraphics[width=0.9\columnwidth]{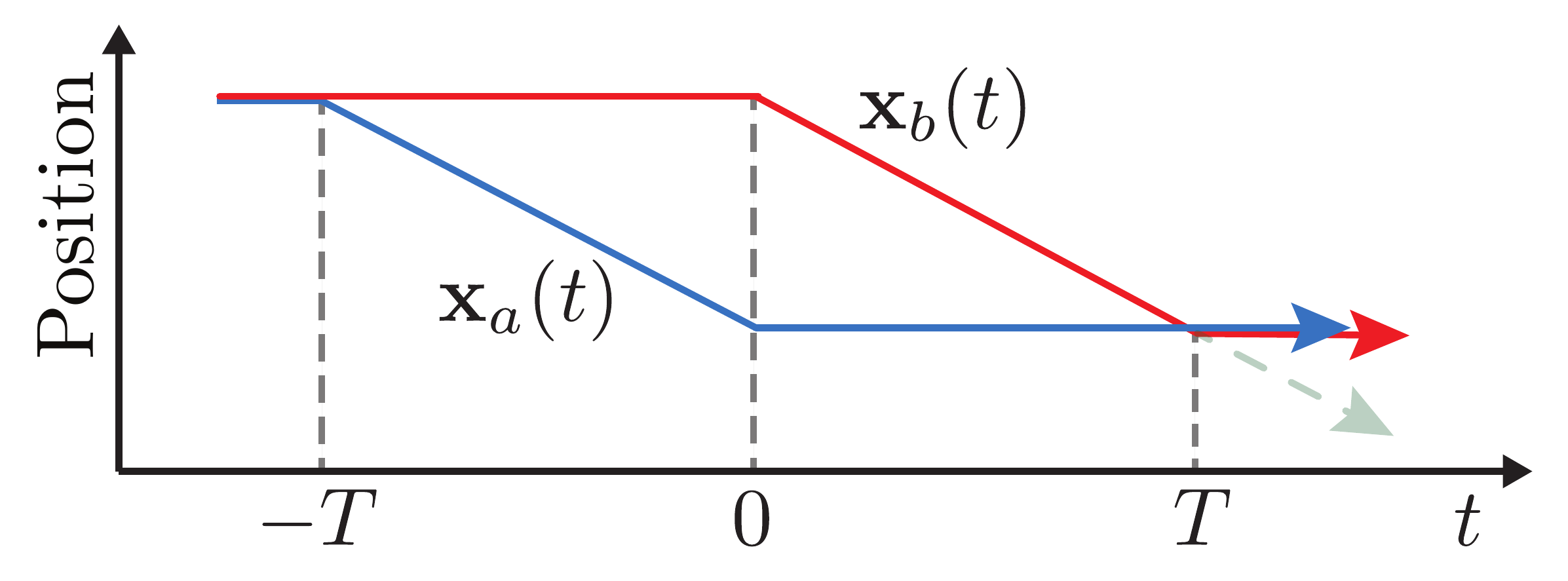}
 \caption{Diagram illustrating the paths in a Mach-Zehnder interferometer.}
 \label{fig:MZpic}
\end{figure}

First consider the case of a $\frac{\pi}{2}-\pi-\frac{\pi}{2}$ Bragg-based Mach-Zehnder
 interferometer, in the absence of rotation. In this case the acceleration on each path is given by 
 
\begin{eqnarray}
    \mathbf{\tilde{a}}_a(t) &=&  \frac{2n\hbar \mathbf{k}}{m}\left[\delta (t+T)-\delta (t)\right] 
    \nonumber
\\  \mathbf{\tilde{a}}_b(t) &=&  \frac{2n\hbar \mathbf{k}}{m}\left[\delta (t)-\delta (t-T)\right] 
\label{eq:MZkicks}
\end{eqnarray}
and it can be seen that these satisfy our time symmetry requirement Eq.~\ref{eq:Vsym}. The space-time area is easy to calculate in this case:\begin{eqnarray}
\mathbf{\mathcal{A}}&=&\frac{2n\hbar \mathbf{k}T^2}{m}
\end{eqnarray}
so the interferometer phase becomes
\begin{eqnarray}
\Delta \Phi &=&  \frac{m\mathbf{g}}{\hbar}\cdot\frac{2n\hbar \mathbf{k}T^2}{m}
\\&=&2n\mathbf{k}\cdot\mathbf{g}T^2
\label{eq:PhaseMZ}
\end{eqnarray}

while the laser phase is 

\begin{eqnarray}
\phi_L&=&\underbrace{\left(n\phi_1-n\phi_2\right)}_{\text{from path 1}}+\underbrace{\left(-n\phi_2+n\phi_3\right)}_{\text{from path 2}}\\
&=&n\left(\phi_1-2\phi_2+\phi_3\right)\,.
\label{eq:LaserMZ}
\end{eqnarray}

With the inclusion of the Coriolis effect, and under the assumption that $\mathbf{g}$ and $\mathbf{k}$ are parallel, the interferometric phase becomes

\begin{eqnarray}
\Delta \Phi &=&2n\mathbf{k}\cdot\left(\mathbf{g}-2 \mathbf{\Omega}\times\mathbf{v}_i\right)T^2.
\label{eq:PhaseMZCoriolis}
\end{eqnarray}

A Mach-Zehnder is sensitive to vibrations according to Eq. (\ref{eq:VibSensCos}). In this case the relative sensitivity to acceleration is given by

\begin{eqnarray}
\mathcal{R}(\omega)&=&\frac{4\sin^2\left(\frac{\omega T}{2}\right)}{(\omega T)^2}
%\label{eq:VibRelMZ}
\end{eqnarray}

which is plotted on Fig. \ref{fig:VibrationSensitivity}, and so the phase shift due to an acceleration $\mathbf{a}=\mathbf{a}_c \cos(\omega t)$ is given by

\begin{align}
\delta\Phi&=\frac{m\mathcal{R}(\omega)}{\hbar}\mathbf{a}_c\cdot\mathcal{A}\\
&=2n\mathbf{k}\cdot\mathbf{a}_c\frac{4\sin^2\left(\frac{\omega T}{2}\right)}{\omega^2}\,.
\end{align}

\subsection{Continuous-Acceleration Bloch (CAB) sequence}

\begin{figure}[h]
\centering{}
 \includegraphics[width=0.9\columnwidth]{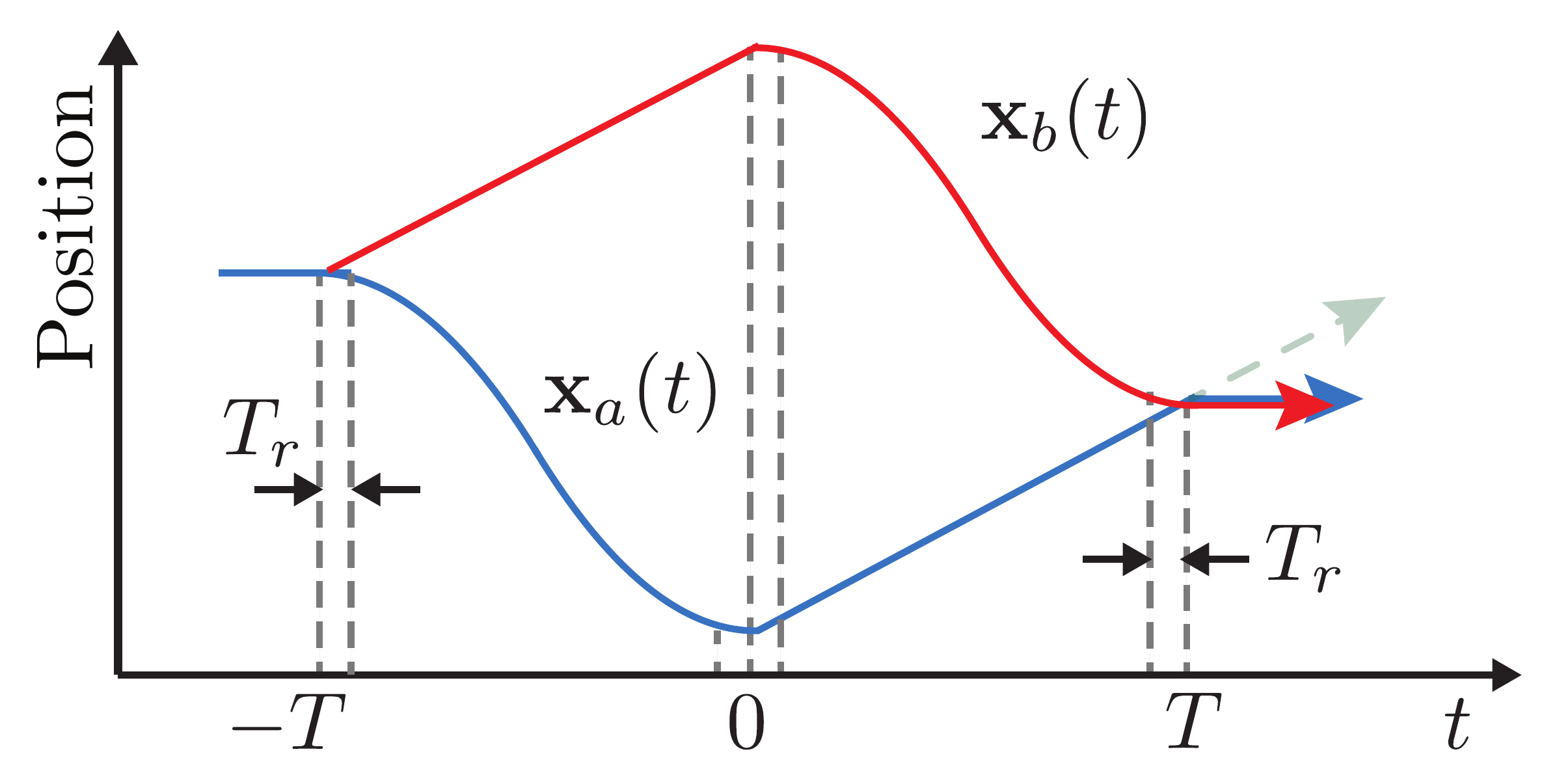}
 \caption{Diagram illustrating the paths in a CAB interferometer sequence. The Bloch lattice is increasing (decreasing) in intensity for a time $T_r$ after (before) each Bragg kick. In the example here, the lattice is not unloaded and reloaded at ${t\approx\pm T/2}$, unlike in the experimental configuration discussed in Refs. \cite{FasterScaling,80hkPRA}.}
 \label{fig:CABpic}
\end{figure}

An extension to the Mach-Zehnder in which the inertial acceleration signal scales as $T^3$ has been demonstrated in Refs. \cite{FasterScaling,80hkPRA}, through the use of Bloch lattice accelerations applied to each arm of the Mach-Zehnder interferometer.
 In this case the acceleration along each path is given by Eq.~(\ref{eq:MZkicks})
 with the addition of a constant Bloch acceleration along one arm at a time, 
 during each half of the interferometer, i.e.
 
\begin{eqnarray}
    \mathbf{\tilde{a}}_a(t) &=& \frac{2n\hbar \mathbf{k}}{m}\left[\delta (t)-\delta (t-T)\right]
              +\frac{2n_{b}\hbar \mathbf{k}}{m\tau_b}\cdot (T_r-T<t<-T_r)\cdot(-1)^{(t>-T/2)}\nonumber
\\  \mathbf{\tilde{a}}_b(t) &=& \underbrace{
\frac{2n\hbar \mathbf{k}}{m}\left[\delta (t+T)-\delta (t)\right]}_{\text{Bragg Kicks}}+\underbrace{
\frac{2n_{b}\hbar \mathbf{k}}{m\tau_b}\cdot(T_r<t<T-T_r)\cdot(-1)^{(t>T/2)} }_{\text{Bloch Oscillations}}
\label{eq:CABkicks}
\end{eqnarray}
where $n_b$ is the number of Bloch oscillations, $\tau_b$ is the period for one Bloch oscillation, and $T_r$ is small time in which the atoms are loaded into the lattice and there is no acceleration. 
In this expression I have used the notation $(x < y)$ to mean a boolean function which is 1 if 
the condition $x<y$ is satisfied and 0 if it is not. The space-time area in this case 
is calculated to be

\begin{eqnarray}
\mathcal{A}&=&\left[\frac{2n\hbar \mathbf{k}T}{m}+\frac{2n_b \hbar \mathbf{k} (T-4T_r)}{2m }\right]\cdot T 
\nonumber
\\&=&\frac{2\hbar \mathbf{k}T^2}{m}\left[n+n_b\left(\frac{1}{2}-\frac{2T_r}{T}\right)\right]
\end{eqnarray}
so the interferometer phase becomes (in the absence of rotation)
\begin{eqnarray}
\Delta \Phi &=&  \frac{m\mathbf{g}}{\hbar}\cdot\frac{2\hbar \mathbf{k}T}{m}\left[n+n_b\left(\frac{1}{2}-\frac{2T_r}{T}\right)\right]
\\&=&2\left[n+n_b\left(\frac{1}{2}-\frac{2T_r}{T}\right)\right]\mathbf{k}\cdot\mathbf{g}T^2
\label{eq:PhaseCAB}
\end{eqnarray}

while the laser phase is 

\begin{eqnarray}
\phi_L&=&\underbrace{\left(n\phi_1-n\phi_2-n_b\phi_{b3}+n_b\phi_{b4}\right)}_{\text{from path 1}}+\underbrace{\left(n_b\phi_{b1}-n_b\phi_{b2}-n\phi_2+n\phi_3\right)}_{\text{from path 2}}\\
&=&n\left(\phi_1-2\phi_2+\phi_3\right)+n_b\left(\phi_{b1}-\phi_{b2}-\phi_{b3}+\phi_{b4}\right)\,.
\label{eq:LaserMZ}
\end{eqnarray}
who implies that an interferometric fringe can be scanned out by changing the phase of the Bloch lattices, as well as the Bragg pulses.

Since the bloch acceleration is constant, we can write the number of bloch 
oscillations as $n_b=\frac{T-4Tr}{2\tau}$, so the interferometric phase in Eq. (\ref{eq:PhaseCAB}) becomes

\begin{eqnarray}
\Delta \Phi &=&  \frac{m\mathbf{g}}{\hbar}\cdot\frac{2\hbar \mathbf{k}T}{m}\left[n+n_b\left(\frac{1}{2}-\frac{2T_r}{T}\right)\right]
\\&=&2\left[n+\frac{T-4T_r}{2\tau}\left(\frac{1}{2}-\frac{2T_r}{T}\right)\right]\mathbf{k}\cdot\mathbf{g}T^2\,
\\&=&2\left[n+\frac{T}{\tau}\left(\frac{1}{2}-\frac{2T_r}{T}\right)^2\right]\mathbf{k}\cdot\mathbf{g}T^2\,.
\end{eqnarray}

and when $T\gg T_r$ we have
 
\begin{eqnarray}
\Delta \Phi &=&2\left[n+\frac{T}{4\tau}\right]\mathbf{k}\cdot\mathbf{g}T^2\,
\\&=&2n\mathbf{k}\cdot\mathbf{g}T^2+\frac{\mathbf{k}\cdot\mathbf{g}T^3}{2\tau}\,
\end{eqnarray}

which shows the $T^3$ sensitivity to acceleration clearly. This can of course also be written as

\begin{eqnarray}
\Delta \Phi &=&\left(2nT^2+\frac{T^3}{2\tau}\right)\mathbf{k}\cdot\mathbf{g}\,
\end{eqnarray}

The inclusion of the Coriolis effect (again under the assumption that $\mathbf{g}\times\mathbf{k}=0$) then changes the interferometric phase to

\begin{eqnarray}
\Delta \Phi &=&\left(2nT^2+\frac{T^3}{2\tau}\right)\mathbf{k}\cdot\left(\mathbf{g}-2 \mathbf{\Omega}\times\mathbf{v}_i\right)\,.
\end{eqnarray}

\begin{SCfigure}
\centering{}
 \includegraphics[width=0.6\columnwidth]{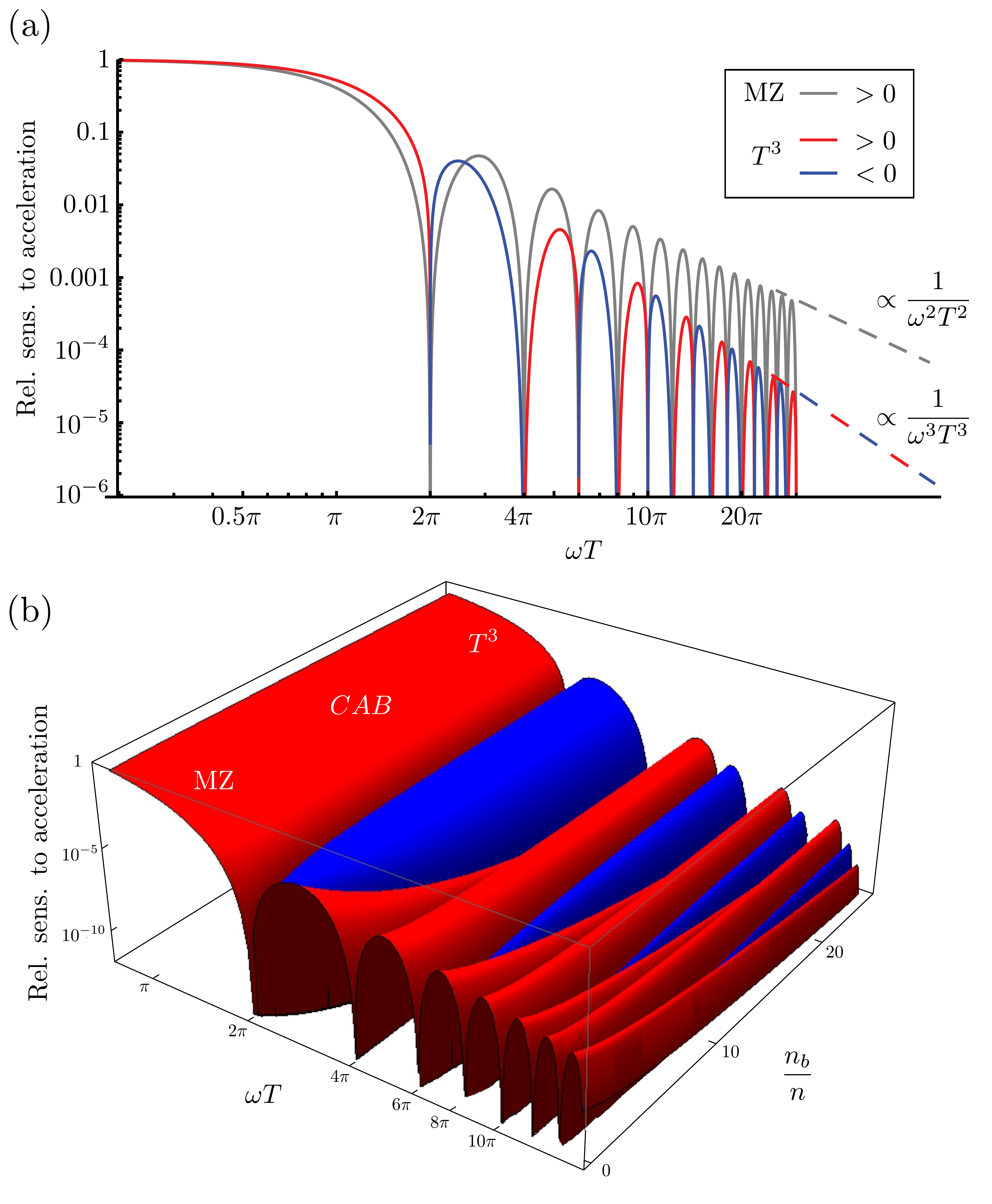}
 \caption{(a) The relative sensitivity $\mathcal{R}$ to an oscillating acceleration with frequency $\omega$ is plotted for the two extreme cases: a Mach-Zehnder configuration, which exhibits a $\omega^2$ roll-off, and the $T^3$ case, which exhibits a $\omega^3$ roll-off. (b) The CAB configuration has an intermediate sensitivity to vibrations, which can look like either of the two extremes depending upon the ratio of Bloch oscillations to Bragg diffraction recoils, $\frac{n_b}{n}$. Note that the CAB configuration has double the number of vibration frequencies to which it is completely insensitive, as compared to either of the extreme cases. }
 \label{fig:VibrationSensitivity}
\end{SCfigure}

The CAB scheme is sensitive to vibrations, again according to Eq. (\ref{eq:VibSensCos}). In the limit of $T\gg T_r$ the relative sensitivity to acceleration is given by

\begin{eqnarray}
\mathcal{R}_{CAB}(\omega)&=&\frac{1}{1+\epsilon}\frac{4\sin^2\left(\frac{\omega T}{2}\right)}{(\omega T)^2}
+\frac{1}{1+\frac{1}{\epsilon}}\frac{64\cos\left(\frac{\omega T}{4}\right)\sin^3\left(\frac{\omega T}{4}\right)}{(\omega T)^3}\\
&=&\frac{1}{1+\epsilon}\mathcal{R}_{MZ}(\omega)
+\frac{1}{1+\frac{1}{\epsilon}}\mathcal{R}_{T^3}(\omega)
\label{eq:VibRelCAB}
\end{eqnarray}

for $\epsilon=\frac{n_b}{2n}$. Thus as $\epsilon$ approaches zero, the noise sensitivity is that of a Mach-Zehnder, whereas when $\epsilon$ is large, the noise sensitivity approaches that of a pure acceleration separation between the arms of the interferometer, i.e. $T^3$ sensitivity. All three cases are illustrated in Figure \ref{fig:VibrationSensitivity}.

The phase shift due to an acceleration $\mathbf{a}=\mathbf{a}_c \cos(\omega t)$ is given by

\begin{align}
\delta\Phi&=\frac{m\mathcal{R}_{CAB}(\omega)}{\hbar}\mathbf{a}_c\cdot\mathcal{A}
\end{align}
which for large $\epsilon$ is that of a $T^3$ interferometer,

\begin{align}
\delta\Phi&=\frac{m\mathcal{R}_{T^3}(\omega)}{\hbar}\mathbf{a}_c\cdot\mathcal{A}\\
&=\frac{64\cos\left(\frac{\omega T}{4}\right)\sin^3\left(\frac{\omega T}{4}\right)}{\omega^3}\frac{\mathbf{k}\cdot\mathbf{g}}{2\tau}\,.
\end{align}

\subsection{Butterfly configuration}
A butterfly configuration is a $\frac{\pi}{2}-\pi-\pi-\frac{\pi}{2}$ Bragg-based interferometer. In this case the acceleration on each path is given by 
 
 \begin{figure}[h]
\centering{}
 \includegraphics[width=0.9\columnwidth]{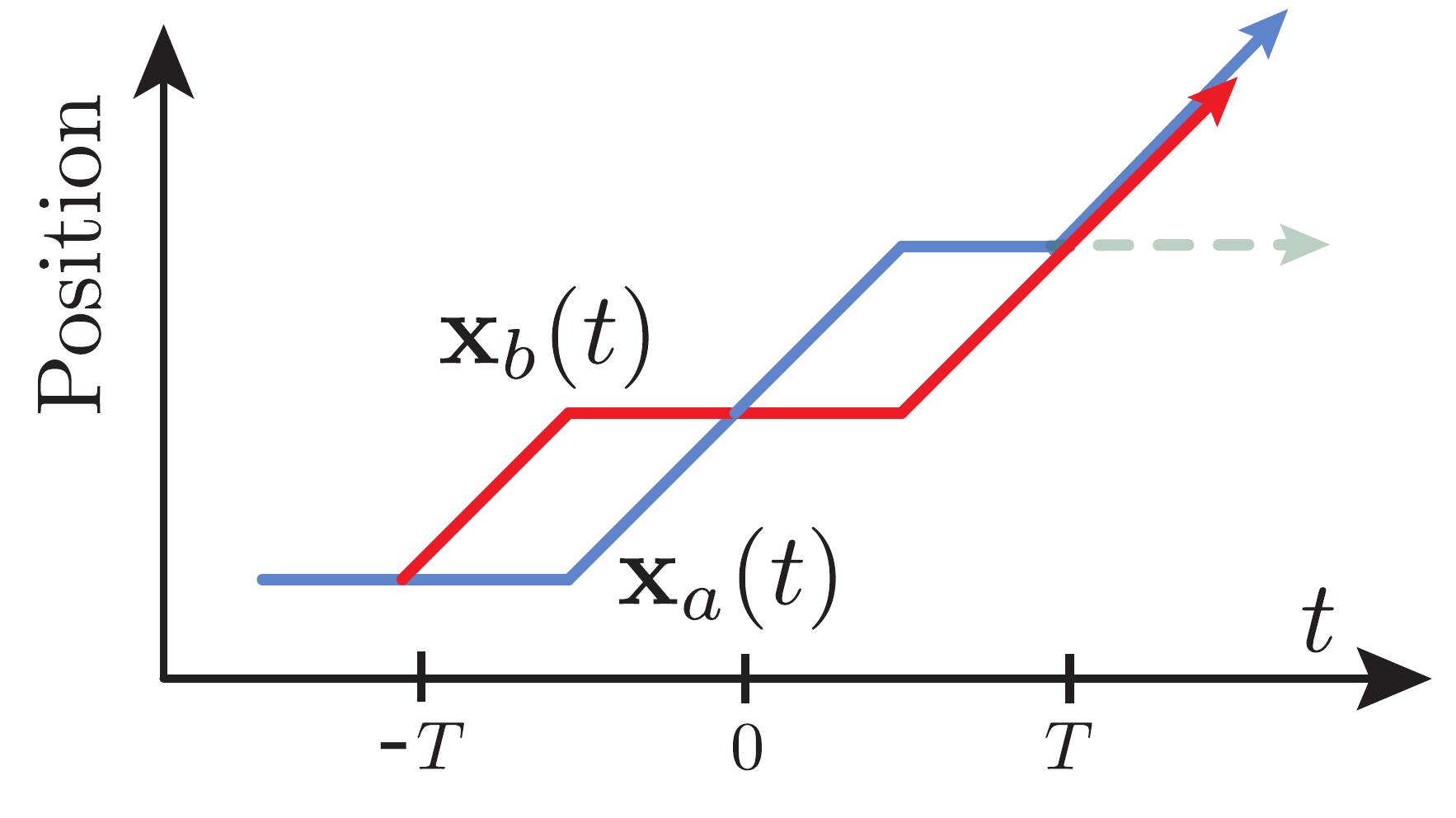}
 \caption{Diagram illustrating the paths in a Butterfly interferometer sequence, which is insensitive to constant accelerations.}
 \label{fig:Butterflypic}
\end{figure}

\begin{eqnarray}
    \mathbf{\tilde{a}}_a(t) &=& \frac{2n\hbar \mathbf{k}}{m}\left[\delta (t+\frac{T}{2})-\delta (t-\frac{T}{2})+\delta(t-T)\right] 
    \nonumber
\\  \mathbf{\tilde{a}}_b(t) &=&  \frac{2n\hbar \mathbf{k}}{m}\left[\delta (t+T)-\delta (t+\frac{T}{2})+\delta (t-\frac{T}{2})\right] 
\label{eq:Butterflykicks}
\end{eqnarray}

\begin{figure}[h]
\centering{}
 \includegraphics[width=0.9\columnwidth]{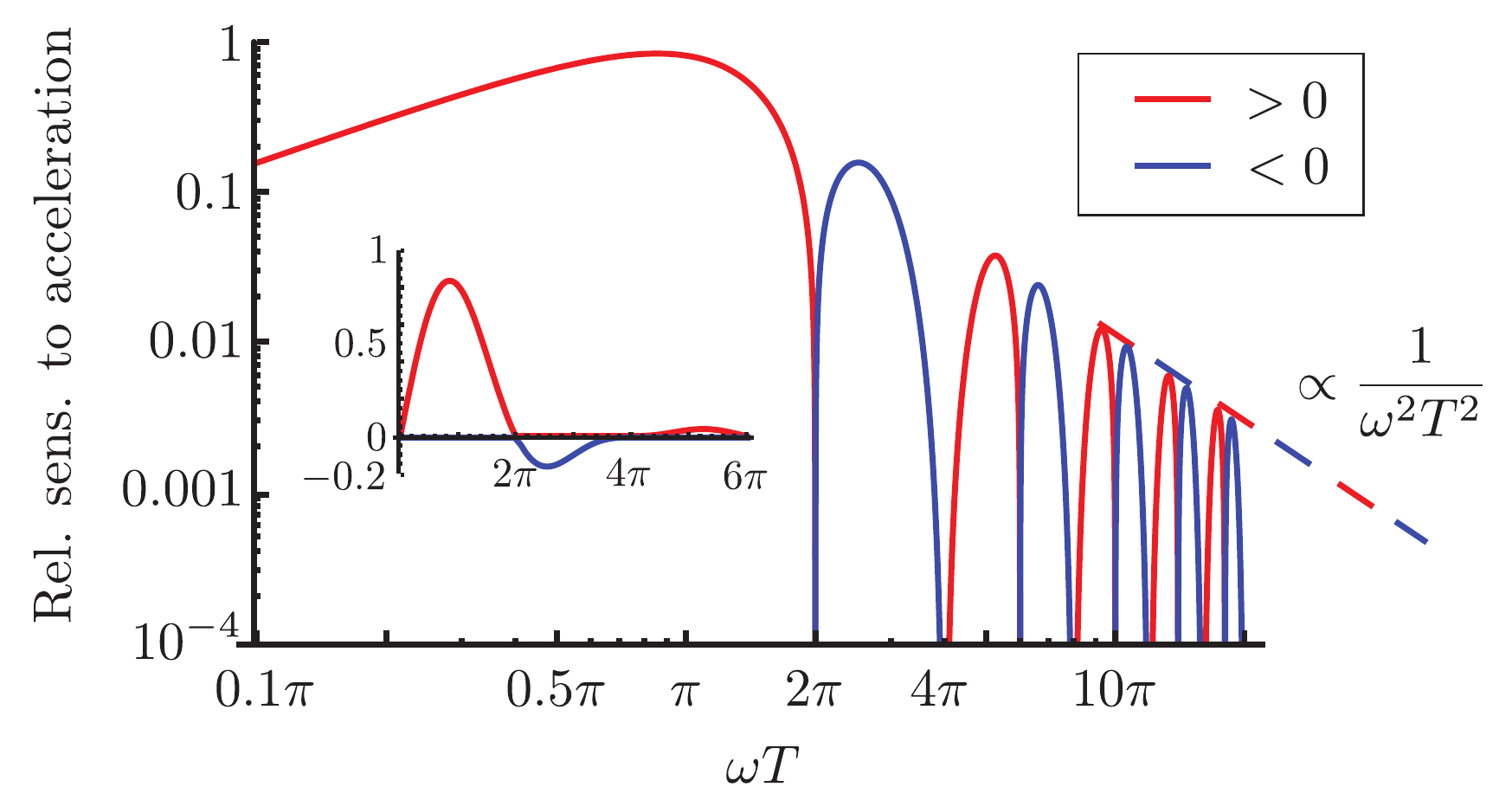}
 \caption{Diagram illustrating $\mathcal{R^*}$, the relative sensitivity to vibrations of the butterfly configuration. This configuration is insensitive to constant acceleration, and at higher frequencies its sensitivity rolls off as $\omega^2$. Inset: Same plot on a linear-linear scale to show that the sensitivity goes to zero for a constant acceleration.}
 \label{fig:Butterflynoise}
\end{figure}

The space-time area is easy to calculate in this case - it is zero, as one parallelogram cancels the other. So this is a constant-acceleration-insensitive configuration, useful for testing the effect of vibration noise in a system. Due to the anti-symmetric nature of this configuration, the sine terms of a vibration now contribute, while the cosine terms do not. Thus the relative acceleration sensitivity becomes

\begin{eqnarray}
\mathcal{R^*}(\omega)&=&\frac{|\mathcal{A}_s(\omega)|}{|\mathcal{A^*}|}\\
&=&\frac{32 \sin ^3\left(\frac{\omega T}{4}\right) \cos \left(\frac{\omega T}{4}\right)}{\omega T^2}\\
\end{eqnarray}
 which is plotted in Fig. \ref{fig:Butterflynoise}. So the phase shift due to an acceleration $\mathbf{a}=\mathbf{a}_s \sin(\omega t)$ is given by
\begin{align}
\delta\Phi&=\frac{m\mathcal{R^*}(\omega)}{\hbar}\mathbf{a}_s\cdot\mathcal{A^*}\\
&=2n\mathbf{k}\cdot\mathbf{a}_s\frac{16\cos\left(\frac{\omega T}{4}\right)\sin^3\left(\frac{\omega T}{4}\right)}{\omega^2}\\
&\approx\frac{n\mathbf{k}\cdot\mathbf{a}_s \omega T^3}{2}
\end{align}
for small $\omega T$.

\subsection{Recoil sensitive interferometers}

Interferometer configurations in which $\Delta\Phi_{\text{kin}}\neq0$ are in general also sensitive to the recoil frequency $\omega_{rec}=\frac{\hbar k^2}{2m}$, since
\begin{align}
\Delta\Phi_{\text{kin}}&=\int_{-T}^{T}\tilde{\omega}_b-\tilde{\omega}_a\,dt\\
&=\omega_{rec}\int_{-T}^{T}\tilde{n}^2_b(t)-\tilde{n}^2_a(t)\,dt
\end{align}
where $\tilde{n}_i(t)$ is the number of photon recoils in the velocity along path $i$ at a given time $t$.

\begin{figure}[h]
\centering{}
 \includegraphics[width=0.9\columnwidth]{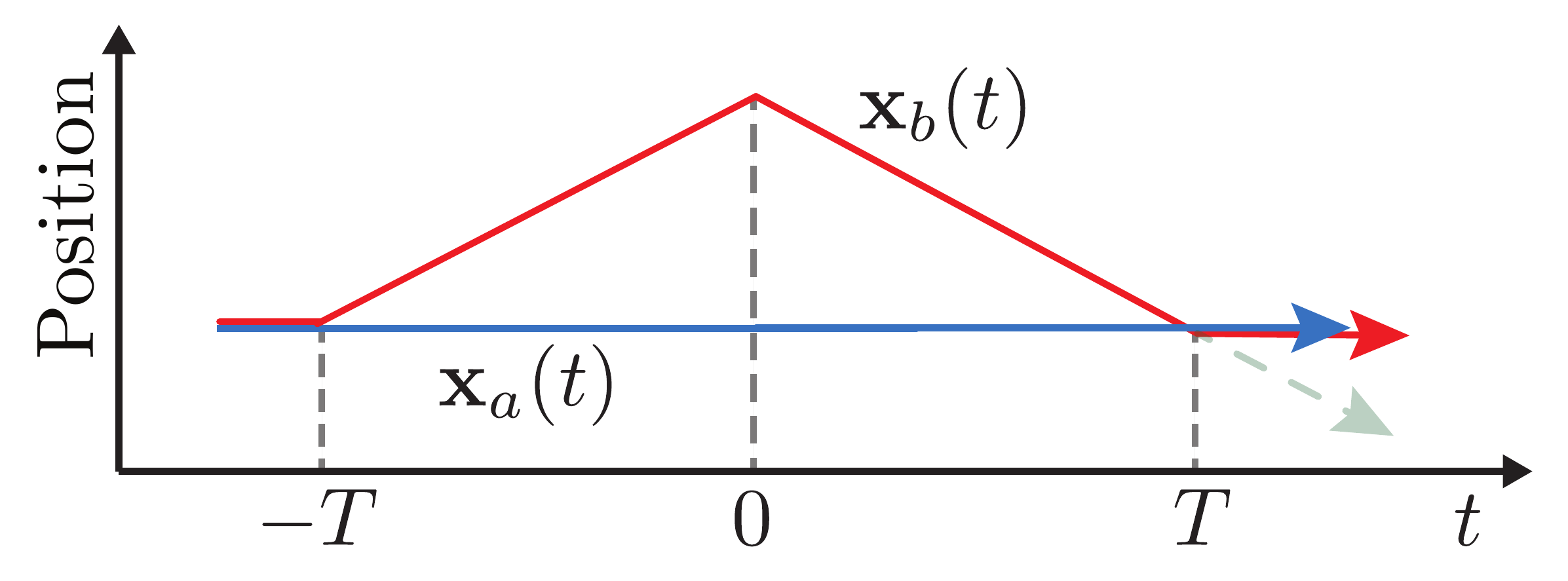}
 \caption{Diagram illustrating the paths in an asymmetric, recoil frequency sensitive interferometer sequence.}
 \label{fig:trianglepic}
\end{figure}

Consider the triangular configuration depicted in Figure \ref{fig:trianglepic}. We assume that the velocity along path $a$ is zero in the inertial frame $\mathbf{\tilde{v}}_a=0$, and the velocity along path $b$ is $\mathbf{\tilde{v}}_b=\pm\mathbf{v}=\pm\frac{2n \hbar \mathbf{k}}{m}$ as shown in Fig.  \ref{fig:trianglepic}. Then the phase is given by

\begin{align}
\Delta\Phi&=\quad\qquad\qquad\Delta\Phi_{\text{kin}}\qquad+\qquad\Delta\Phi_{\text{inertial}}\\
&=\omega_{rec}\int_{-T}^{T}\tilde{n}^2_b(t)-\tilde{n}^2_a(t)\,dt+\frac{m}{\hbar}\int_{-T}^{T} \mathbf{g}\cdot\Delta\mathbf{\tilde{x}}\,dt\\
&=\quad\qquad\qquad8n^2\omega_{rec}T\qquad-\qquad2n\mathbf{k}\cdot\mathbf{g}T^2\\
\end{align}

If we instead choose a constant acceleration separation between the two states, i.e. the velocity goes as 
\begin{displaymath}
\mathbf{\tilde{v}}(t)= \left\{
     \begin{array}{lr}
       (t+T)\,\mathbf{a}\qquad\text{for}\qquad t<-\frac{T}{2}\\
       -(t)\,\mathbf{a}\qquad\text{for}\qquad -\frac{T}{2}<t<\frac{T}{2}\\
       (t-T)\,\mathbf{a}\qquad\text{for}\qquad t>\frac{T}{2}
     \end{array}
   \right.
\end{displaymath}

then we have the interferometric phase\begin{eqnarray}
\Delta \Phi &=&\frac{m}{\hbar}\left(\int_{-T}^{T}|\mathbf{v}|^2\, ds+\mathbf{g}\cdot\mathcal{A}\right)\\
&=&\frac{m}{\hbar}\left(4\int_{0}^{T/2}|\mathbf{a}|^2\, s^2ds-\frac{\mathbf{g}\cdot\mathbf{a}}{4}\right)T^3\\
&=&\frac{m}{\hbar}\left(\frac{|\mathbf{a}|^2}{6}-\frac{\mathbf{g}\cdot\mathbf{a}}{4}\right)T^3\\
\end{eqnarray}
and so our sensitivity to the recoil frequency (which is now buried in $|\mathbf{a}|^2$) goes as $T^3$. For instance if the constant acceleration is due to $n_b\,$ $2\hbar \mathbf{k}$-Bloch oscillations over each time $T/2$, then $\mathbf{a}=\frac{4n_b\hbar\mathbf{k}}{mT}$, and the phase becomes
\begin{eqnarray}
\Delta \Phi &=&\frac{8}{3}n_b^2 \omega_{rec} T-n_b\mathbf{k}\cdot\mathbf{g}T^2\\
&=&\left(\frac{2}{3}\frac{\omega_{rec}}{\tau_b^2}-\frac{\mathbf{k}\cdot\mathbf{g}}{2\tau_b}\right)T^3
\end{eqnarray}
where on the last line we have substituted $n_b=\frac{T}{2\tau_b}$ for a constant time $\tau_b$ for each Bloch oscillation.
As we keep higher derivatives of position (let's say the $p$-th derivative) constant, the space-time area will increase as $T^{p+1}$, whereas the recoil-dependant term will increase as $T^{2p-1}$.

To build an acceleration-insensitive configuration you can put two of these back to back in such a way as to cancel the acceleration signal as in Ref. \cite{ContrastInterferometry}. This is done by adding all 4 paths together such that the central two add constructively. It is also possible to put them together in such a way as to cancel the recoil phase shift instead, and this is equivalent to a Mach-Zehnder with twice the momentum splitting \cite{PhysRevA.81.013617}. This is done by adding all 4 paths together such that (conceptually at least) the central two interfere destructively and cancel out.

\section{Effect of $s$-wave interactions}
\label{sec:InTrapAITheory-PhaseShift}

So far we have studied interferometry while treating each atom as an individual, with there being no interactions between them, i.e. no collisions. In various situations it is necessary to address the effect of these interactions. Such situations include free-space interferometers with high phase-space density, waveguides interferometers with high phase-space density, and in-trap interferometers such as via the magnetic internal states of an atom. The generalisation of the Schr\"odinger equation to simplify the many-body problem of a BEC is the Gross-Pitaevski equation:

\begin{align}
i\hbar\frac{\partial\psi}{\partial t}=-\frac{\hbar^2}{2m}\nabla^2\psi+V(\mathbf{x})\psi+\frac{4\pi \hbar^2 a_s}{m}\left|\psi\right|^2\psi
\label{GPE}
\end{align}
 where the additional term involves the $s$-wave scattering length $a_s$, and the density $n(\mathbf{x})=\left|\psi\right|^2$ at each position in space. The wave-function $\psi$ has been replaced by the analogous order parameter, as it now represents a kind of average over what is in reality a large and complicated many-body tensor-product wave-function.

Assuming no dynamic effects of such a perturbation to the Hamiltonian, the additional energy density of $\mathcal{E}_{ii}=\frac{4\pi\hbar^2 a_s}{m} n(\mathbf{x})^2$ would add a positionally-dependent phase shift of

\begin{align}
\Delta\Phi_{\text{int}}=\frac{m}{\hbar}\int_{-T}^{T}\Delta\mathcal{E} dt
\end{align}

However in many cases where interactions become relevant in atom interferometry, this energy difference also causes dynamics to occur in the system. In these cases numerical simulation of the interferometer through Eq. (\ref{GPE}) becomes appropriate.

\bibliography{STABibliography.bib}
\end{document}